\RequirePackage{fix-cm}
\RequirePackage{fixltx2e}
\documentclass[12pt,english]{article}
\usepackage[T1]{fontenc}
\usepackage[latin9]{inputenc}
\usepackage{geometry}
\usepackage{amsmath}
\usepackage{amssymb}

\makeatletter
\@ifundefined{date}{}{\date{}}
\usepackage{amsfonts,amsthm,amsmath,amssymb,upgreek}
\usepackage{hyperref}
\usepackage{cite}
\usepackage{units}
\usepackage{color}
\usepackage[export]{adjustbox}
\usepackage{mathtools}
\usepackage{physics}
\usepackage{mciteplus}

\numberwithin{equation}{section}

\mciteSetMidEndSepPunct{}{\ifmciteBstWouldAddEndPunct.\else\fi}{\relax}
\renewcommand\[{\begin{equation}} 
\renewcommand\]{\end{equation}}
\renewenvironment{align*}{\align}{\endalign}

\renewcommand{\title}[1]{\vbox{\center\bf{\Large{#1}}}\vspace{5mm}}
\renewcommand{\author}[1]{\vbox{\center#1}\vspace{5mm}}
\newcommand{\address}[1]{\vbox{\center\em#1}}
\newcommand{\email}[1]{\vbox{\center\tt#1}\vspace{5mm}}

\makeatother

\usepackage{babel}
\begin{document}
\begin{titlepage}

\begin{center} 
\hfill \\ 
\hfill \\ 
\vskip .5cm

\title{SYK Correlators for All Energies}

\author{Alexandre Streicher$^{abcd}$} 

\address{$^{a}$ Institute for Advanced Study,\\ Princeton, NJ 08540, USA}

\vspace{10pt}

\address{$^{b}$ Perimeter Institute for Theoretical Physics\\ Waterloo, ON N2L 2Y5, Canada}

\vspace{10pt}

\address{$^{c}$ Stanford Institute for Theoretical Physics, Stanford University,\\ Stanford, CA 94305, USA}

\vspace{10pt}

\address{$^{d}$ Department of Physics, University of California,\\ Santa Barbara, CA 93106, USA}

\email{streicher@ias.edu, astreicher1@perimeterinstitute.ca} 

\end{center}
\begin{abstract}
The Sachdev-Ye-Kitaev (SYK) model, a theory of $N$ Majorana fermions
with $q$-body interactions, becomes in the large $q$ limit a conformally-broken
Liouville field theory. Taking this limit preserves many interesting
properties of the model, yet makes the theory as a whole much more
tractable. Accordingly, we produce novel expressions for the two and
four-point correlators at arbitrary temperature and find the surprising
result they take a universal closed form. We note that these expressions
correctly match onto and interpolate between previously-obtained low-energy
results and simple high-energy perturbative checks. We find that the
time-ordered four-point correlators are always determined by finite
temperature OPEs into the identity and Hamiltonian, while the out-of-time-order
four-point correlators remain nontrivial and always scramble. This
has only been established in the conformal limit, so to find that
it holds for large $q$ at all temperatures/couplings is a nontrivial
result. Finally, we determine the system's thermalization and scrambling
rates and find that they always agree, regardless of temperature.
This adds to the increasing body of evidence that there exists simple
structures in large $N$ internal dynamics, such as those formed by
SYK's epidemic operator growth. 
\end{abstract}
\end{titlepage}
\tableofcontents{}

\section{Introduction}

The Sachdev-Ye-Kitaev model is one of the only non-integrable yet
solvable models of many-body physics, containing $N$ flavors of Majorana
fermions all coupled together in $q$-body interactions \cite{Sachdev:1992fk,Kitaev:2014t1}.
The model is made tractable by taking an large $N$ t'Hooft limit,
which results in the theory becoming dominated by the infinite ``melonic''
subset of planar diagrams, rather than the intractable set of all
planar diagrams often found in matrix models \cite{Klebanov_2017}.
In such a large $N$ limit, the low-energy physics was found to saturate
a chaos bound on the decorrelation often out-of-time-order (OTO) four-point
correlators \cite{Kitaev:2014t2,Maldacena:2016hyu,Maldacena:2015waa},
hinting at a rich underlying structure for this model containing only
internal degrees of freedom. 

This physics was shown to arise from the near-conformal low-energy
behavior of the model, where fluctuations are universally governed
by a Schwarzian action \cite{Maldacena:2016upp,Kitaev2018,Maldacena:2016hyu}.
Such a one-dimensional action naturally arises by recasting $2D$
Jackiw-Teitelboim gravity into its $1D$ boundary holographic action
\cite{Maldacena:2016upp}. Alternatively, one may take a $1\ll q\ll N$
limit of the SYK model to arrive at an effective description in terms
of broken $1D$ Liouville field theory, whose action is only a field
redefinition away from that of the Schwarzian \cite{Maldacena:2018lmt}.
Indeed, after taking the t'Hooft limit, this large $q$ limit makes
the model even more tractable, as the physics of broken Liouville
theory is much easier to analyze than generic broken conformal physics. 

Recently this limit was used to find details of $SYK$ operator evolution,
where a microscopic wavefunction amongst the internal degrees of freedom
spread in a manner matching that of a primary operator ``falling''
along higher weight operators/descendants (or equivalently, how such
a boundary operator propagates deeper into the bulk), when evolved
using one of the non-compact time-like generators of the $1D$ conformal
group \cite{Qi:2018bje,Brown:2018kvn}. Furthermore, they were able
to determine that this behavior was universal in the large $q$ limit,
regardless of temperature/energy, so the graph-theoretic epidemic
intuition from infinite temperature evolution remained relevant \cite{Roberts:2018mnp}.

In Appendix \ref{sec:Twisting-Correlators} we review how two-point
functions ``twisted'' to satisfy Euclidean-time bi-local boundary
conditions generate higher correlators. Of course, it is extremely
difficult to solve for such an object in a generic theory, as even
determining the regular two-point function may be impossible. However,
by utilizing the solvability of the broken Liouville large $q$ limit
of the SYK model, we are able to find such a ``twisted'' two-point
function. Turning the crank, we obtain the large $q$ two-point correlator
as well as the time-ordered and out-of-time-ordered four-point correlators
for arbitrary energies and couplings. We confirm that these expressions
correctly match onto both weak-coupling/high-temperature perturbative
results and previously-derived strong-coupling/low-temperature expressions
\cite{Maldacena:2016hyu}. We have thus produced the first set of
resumed expressions for four-point correlators at arbitrary energies.

There exist several interesting phenomenon as one varies the energy
and/or coupling. First we note that at all energies, the time-ordered
four-point function is entirely determined by finite temperature OPEs
into the identity and Hamiltonian, which has only previously been
shown to hold in the near-conformal limit \cite{Maldacena:2016hyu}.
That is, for all intents and purposes, when there are no other operators
between two flavor-averaged fermions on the thermal circle, one has
that to first nontrivial order in $N$
\begin{align*}
\hat{G}_{12}\equiv\frac{1}{N}\sum_{i=1}^{N}\hat{\psi}_{i}\left(\tau_{1}\right)\hat{\psi}_{i}\left(\tau_{2}\right) & =\left\langle \hat{G}_{12}\right\rangle _{\beta}\hat{\mathbb{I}}+\frac{\left\langle \delta\hat{H}\hat{G}_{12}\right\rangle _{\beta}}{\left\langle (\delta\hat{H})^{2}\right\rangle _{\beta}}\delta\hat{H}
\end{align*}
where $\delta\hat{H}\equiv\hat{H}-\left\langle \hat{H}\right\rangle _{\beta}$
is the Gram-Schmidt-processed Hamiltonian. Such a simple expansion
does not hold for the OTO four-point correlators, which signals that
the large $q$ limit does not just reduce the theory to a free theory,
but rather preserves interesting interactions. In fact, we find that
all of the OTO four-point correlators scramble, regardless of the
energy, with the same rate as that of thermalization
\[
\lambda_{L}\sim\frac{2}{t_{thermal}}\sim\frac{2\pi}{\beta}v\qquad\pi v=\beta\mathcal{J}\cos\frac{\pi v}{2}
\]
where $v=v\left(\beta\mathcal{J}\right)$ is determined by the energy
and coupling of the system \cite{Maldacena:2016hyu}. Of course, the
two timescales are parametrically separated by $\log N$, but it is
still interesting to note that thermalization and scrambling occur
at the same rate for all energies and couplings and not just in the
near-conformal limit. The matching of these two rates has only been
observed in systems such as large $c$ $2D$ $CFT$s where the temperature
controls all rates \cite{Roberts:2014ifa}.

\section{Large $q$ SYK\label{sec:SYK-Model}}

The SYK model features $q$-local interactions with independently
random couplings, where each of the couplings is normal distributed
\begin{equation}
H=i^{q/2}\sum_{1\leq i_{1}...\leq i_{q}\leq N}J_{i_{1}...i_{q}}\psi_{i_{1}}...\psi_{i_{q}}\qquad\left\langle J_{i_{1}...i_{q}}^{2}\right\rangle =\frac{J^{2}}{\binom{N-1}{q-1}}=\frac{\mathcal{J}^{2}}{2q\binom{N-1}{q-1}}\qquad\left\{ \psi_{i},\psi_{j}\right\} =2\delta_{ij}\label{eq:SYK Model}
\end{equation}
At large $N$, the two-point function satisfies the saddle-point equations
\begin{align}
\left[G\right]^{-1} & =[G_{0}]^{-1}-\left[\Sigma\right]\qquad\Sigma\left(\tau_{1},\tau_{2}\right)=\frac{\mathcal{J}^{2}}{2q}\left(G\left(\tau_{1},\tau_{2}\right)\right)^{q-1}\label{eq:Orig Sad-Pt}
\end{align}
where bracketed terms are Matsubara frequency matrices. One should
note that since our fermions square to one, $\left[G_{0}\right]^{-1}=-i\omega/2$
rather than typical $-i\omega$.

\subsection{Large $q$ Approximation}

When generating diagrams using the Schwinger-Dyson equations, note
that diagrams where melons are inserted into melons receive a combinatorial
$q$ enhancement, as there are many rungs upon which one may insert
(hence the need for a $q^{-1}$ factor in the self-energy to keep
everything finite). However, diagrams where melons are simply threaded
together do not receive this enhancement \cite{Gross:2016kjj}. Thus,
at large $q$ only the former dominate, which can be captured by the
following truncation of the Schwinger-Dyson expansion:
\begin{align*}
\left[G\right] & =\left[G_{0}\right]+\left[G_{0}\right]\left[\Sigma\right]\left[G_{0}\right]\qquad\Sigma=\frac{\mathcal{J}^{2}}{2q}G^{q-1}
\end{align*}
Combing the equations together and Fourier transforming, one obtains
\[
\partial_{\theta_{1}}\partial_{\theta_{2}}\left(G-G_{0}\right)=-\frac{2}{q}\left(\frac{\beta\mathcal{J}}{2\pi}\right)^{2}G^{q-1}
\]
The role played by $G_{0}$ in this equation is to require that $G\rightarrow G_{0}$
as $\theta_{12}^{-}$ goes to integer multiples of $2\pi$. If we
accordingly re-express $G=G_{0}e^{\sigma/q}$, then we obtain Liouville's
equation 
\begin{align}
\partial_{\tau_{1}}\partial_{\tau_{2}}\sigma & =-2\left(\frac{\beta\mathcal{J}}{2\pi}\right)^{2}e^{\sigma}+\mathcal{O}\left(1/q\right)\label{eq:Liouville}
\end{align}
where the field $\sigma$ is expected to be periodic in both of its
arguments, as well as have kinks when $\theta_{12}$ approaches integer
multiples of $2\pi$. 

Rather than calculate the original SYK Green's function $G$, we would
like to figure out how to solve Liouville's equation for two-point
function $\mathcal{G}$ obeying the the constraint
\[
\lim_{\theta_{1}/\theta_{2}\rightarrow\theta_{3}^{+}}\left(\begin{array}{c}
\mathcal{G}\left(\theta_{1},\theta_{2}\right)\\
\mathcal{G}\left(2\pi-\theta_{1},\theta_{2}\right)
\end{array}\right)=\left(\begin{array}{cc}
\cosh\left(\mu\right) & -\sinh\left(\mu\right)\\
-\sinh\left(\mu\right) & \cosh\left(\mu\right)
\end{array}\right)\lim_{\theta_{1}/\theta_{2}\rightarrow\theta_{3}^{-}}\left(\begin{array}{c}
\mathcal{G}\left(\theta_{1},\theta_{2}\right)\\
\mathcal{G}\left(2\pi-\theta_{1},\theta_{2}\right)
\end{array}\right)
\]
In the large $q$ limit, the twist conditions decouple and become
\begin{align}
\lim_{\theta_{1}/\theta_{2}\rightarrow\theta_{3}^{+}}\sigma & =\lim_{\theta_{1}/\theta_{2}\rightarrow\theta_{3}^{-}}\sigma-q\nu+\mathcal{O}\left(q^{-1}\right)\label{eq:Large q Bdry Con}
\end{align}
where $\nu=\mu G\left(\theta_{34}^{-}\right)$.

\subsection{Twisted Two-Point Function}

Due to the presence of the operator insertions at $\theta_{3}$ and
$-\theta_{3}$ (a.k.a. $2\pi-\theta_{3}$) and the point reflection
symmetry $\left(\theta_{12}^{-},\theta_{12}^{+}\right)\rightarrow\left(2\pi-\theta_{12}^{-},2\pi-\theta_{12}^{+}\right)$
(\ref{eq:Plus Reflection}), we need only solve for $\mathcal{G}$
in three regions of the domain $0<\theta_{12}^{\pm}<2\pi$. First,
we have the uncrossed twisted correlators
\begin{align*}
\mathcal{G} & =\begin{cases}
G_{0}\left(\frac{\sin\gamma_{1}}{\sin\left(v_{1}\theta_{12}+\gamma_{1}\right)}\right)^{2/q} & \theta_{1}>\theta_{2}>\theta_{3}\\
G_{0}\left(\frac{\sin\gamma_{2}}{\sin\left(v_{2}\theta_{12}+\gamma_{2}\right)}\right)^{2/q} & \theta_{3}>\theta_{1}>\theta_{2}
\end{cases}
\end{align*}
and the crossed twisted correlator
\begin{align}
\mathcal{G} & =G_{0}e^{-\nu}\left(\frac{\sin\gamma_{1}\sin\gamma_{2}}{e^{-q\nu}\sin\left(v_{1}\theta_{13}^{-}\right)\sin\left(v_{2}\theta_{23}^{-}\right)-\sin\left(v_{1}\theta_{13}^{-}+\gamma_{1}\right)\sin\left(v_{2}\theta_{23}^{-}-\gamma_{2}\right)}\right)^{2/q}\quad\theta_{1}>\theta_{3}>\theta_{2}\label{eq:Crossed Config}
\end{align}
Furthermore, Liouville's equation (\ref{eq:Liouville}) constrains
the parameters
\begin{align}
\pi v_{i} & =\beta\mathcal{J}\sin\gamma_{i}\label{eq:Constraint 1}
\end{align}
while the point reflection symmetry $\left(\theta_{12}^{-},\theta_{12}^{+}\right)\rightarrow\left(2\pi-\theta_{12}^{-},2\pi-\theta_{12}^{+}\right)$
(\ref{eq:Plus Reflection}) leads to more constraints
\begin{align}
e^{-q\nu}\sin\left(\frac{v_{1}}{2}\left(\pi-\theta_{3}\right)\pm\frac{v_{2}}{2}\theta_{3}\right) & =\sin\left(\frac{v_{1}}{2}\left(\pi-\theta_{3}\right)\pm\frac{v_{2}}{2}\theta_{3}+\gamma_{1}\pm\gamma_{2}\right)\label{eq:Constraint 2}
\end{align}
Given particular values of $\theta_{3}$, $\beta\mathcal{J}$, and
$\mu$ these four constraints (\ref{eq:Constraint 1}) (\ref{eq:Constraint 2})
uniquely determine $v_{1/2}$ and $\gamma_{1/2}$. Furthermore, when
$\nu=\mu=0$, we have that $v_{1}=v_{2}=v$, $\gamma_{1}=\gamma_{2}=\frac{\pi}{2}\left(1-v\right)$,
and that
\begin{equation}
\pi v=\beta\mathcal{J}\cos\frac{\pi v}{2}\label{eq:determine v}
\end{equation}

\subsection{Two-Point Correlators at All Energies}

By taking the $\mu\rightarrow0$ limit of our twisted two-point function,
we may recover the previously-derived large $q$ expression for the
two-point function at arbitrary temperature \cite{Maldacena:2016hyu}
in the domain $0<\theta_{12}^{\pm}<2\pi$ (with all other values determined
by symmetries)
\[
G=G_{0}\left(\frac{\cos\left(\pi v/2\right)}{\cos\left(v\left(\frac{\pi}{2}-\theta_{12}^{-}\right)\right)}\right)^{2/q}=G_{0}\left(\frac{\pi v}{\beta\mathcal{J}\cos\left(v\left(\frac{\pi}{2}-\theta_{12}^{-}\right)\right)}\right)^{2/q}
\]
We see then that the thermal timescale - the timescale of variation
for the two-point correlator - is not given by what is expected for
either a perturbative theory ($t_{thermal}\sim\mathcal{J}^{-1}$)
nor that of a conformal theory or a theory lacking quasiparticles
($t_{thermal}\sim\beta$). Instead, we find that the thermalization
timescale interpolates between these two behaviors as it is given
by
\[
t_{thermal}\sim\frac{\beta}{v}\sim\frac{1}{\mathcal{J}}\left(1+\frac{\left(\mathcal{\beta}\mathcal{J}\right)^{2}}{8}+\mathcal{O}\left(\left(\beta\mathcal{J}\right)^{4}\right)\right)\sim\beta\left(1+\frac{2}{\beta\mathcal{J}}+\mathcal{O}\left(\left(\beta\mathcal{J}\right)^{-2}\right)\right)
\]
in the perturbative/high-energy and strong-coupling/low-energy regime
respectively. Furthermore, from the equation for $v$ (\ref{eq:determine v}),
we see that the thermalization scale is strictly bounded above by
$\beta$. We conclude no matter how large the coupling is, a conformal
system will always thermalize faster.

\section{Four-Point Correlators at All Energies}

We now have everything we need to obtain the four-point function.
Defining the disconnected and connected four-point correlators $F\equiv F_{d}+\frac{\mathcal{F}}{N}$,
we find that the connected four-point correlator is generated by the
twisted two-point correlator
\[
2\lim_{\nu\rightarrow0}\partial_{\nu}\ln\mathcal{G}\left(\theta_{1},\theta_{2}\right)=\frac{\mathcal{F}\left(\theta_{1},\theta_{2},\theta_{3},-\theta_{3}\right)}{G\left(\theta_{12}^{-}\right)G\left(\theta_{34}^{-}\right)}=\frac{\mathcal{F}\left(\theta_{1},\theta_{2},\theta_{3},-\theta_{3}\right)}{F_{d}}
\]

We remind the reader that we use translation invariance to relate
all four-point correlators to those of the form $F\left(\theta_{1},\theta_{2},\theta_{3},-\theta_{3}\right)$,
followed by the use of anti-periodicity, swap symmetry, and $C\left(P\right)T$
invariance (see Appendix \ref{sec:Fundamental-Domain-of}) to relate
all such correlators to those with $0<\theta_{12}^{\pm}<2\pi$ and
$0<\theta_{3}<\pi$ (a.k.a. $\theta_{34}^{+}=0$ and $0<\theta_{34}^{-}<2\pi$).
In addition, we see that due to the point reflection symmetry $\left(\theta_{12}^{-},\theta_{12}^{+}\right)\rightarrow\left(2\pi-\theta_{12}^{-},2\pi-\theta_{12}^{+}\right)$,
it is sufficient to determine the correlators in only one half of
the above domain. Accordingly, out of the four possible uncrossed
configurations in our domain $\theta_{1}>\theta_{2}>\theta_{3}$,
$\theta_{3}>\theta_{1}>\theta_{2}$, $\theta_{1}>\theta_{2}>-\theta_{3}$,
and $2\pi-\theta_{3}>\theta_{1}>\theta_{2}$, we need only solve for
those satisfying $\theta_{1}>\theta_{2}>\theta_{3}$ and $\theta_{3}>\theta_{1}>\theta_{2}$.
In conclusion, the four-point correlators of the form $F\left(\theta_{1},\theta_{2},\theta_{3},-\theta_{3}\right)$
with $0<\theta_{12}^{\pm}<2\pi$ and $0<\theta_{3}<\pi$, fall into
either crossed or uncrossed configurations, and completely determine
the four-point correlators for all possible times.

\subsection{Time-Ordered Correlators and Finite Temperature OPEs}

If we define
\begin{align*}
f\left(\theta\right) & \equiv\left(1-\left(\frac{v}{2}\theta+\cot\frac{\pi v}{2}\right)\tan\left(\frac{v}{2}\left(\pi-\theta\right)\right)\right)
\end{align*}
then the ratio of the time-ordered connected and disconnected four-point
correlators can be expressed as
\begin{align}
\frac{\mathcal{F}}{F_{d}} & =\frac{2\tan\frac{\pi v}{2}}{\frac{\pi v}{2}+\cot\frac{\pi v}{2}}f\left(\theta_{12}^{-}\right)\times\begin{cases}
f\left(\theta_{34}^{-}\right) & \theta_{1}>\theta_{2}>\theta_{3}\\
f\left(2\pi-\theta_{34}^{-}\right) & \theta_{3}>\theta_{1}>\theta_{2}
\end{cases}\label{eq:uncrossed OTOC}
\end{align}
where $v=v\left(\beta\mathcal{J}\right)$ is determined from (\ref{eq:determine v}). 

Let us now restrict attention to the configuration $\theta_{1}>\theta_{2}>\theta_{3}$.
Then, by defining the bi-local flavor-averaged operators $\hat{G}_{ij}\equiv\hat{G}\left(\theta_{i},\theta_{j}\right)\equiv\frac{1}{N}\sum_{k=1}^{N}\hat{\psi}_{k}\left(\tau_{i}\right)\hat{\psi}_{k}\left(\tau_{j}\right)$,
we may re-express the uncrossed four-point function (\ref{eq:uncrossed OTOC})
as arising from a sum of products of three-point correlators
\begin{align}
F & =G\left(\theta_{12}^{-}\right)G\left(\theta_{34}^{-}\right)+G\left(\theta_{12}^{-}\right)G\left(\theta_{34}^{-}\right)\frac{2\tan\frac{\pi v}{2}}{\frac{\pi v}{2}+\cot\frac{\pi v}{2}}\frac{f\left(\theta_{12}^{-}\right)f\left(\theta_{34}^{-}\right)}{N}\nonumber \\
F & =\left\langle \hat{G}_{12}\hat{G}_{34}\right\rangle =\left\langle \hat{G}_{12}\right\rangle \left\langle \hat{G}_{34}\right\rangle +\frac{\left\langle \delta\hat{H}\hat{G}_{12}\right\rangle \left\langle \delta\hat{H}\hat{G}_{34}\right\rangle }{\left\langle (\delta\hat{H})^{2}\right\rangle }\label{eq:OPE-like relation}
\end{align}
where $\delta\hat{H}\equiv\hat{H}-\left\langle \hat{H}\right\rangle $
and the expressions for $\left\langle \delta\hat{H}\hat{G}_{12}\right\rangle $
may be obtained by writing $G_{12}$ in terms of $\tau$ variables
and then taking derivatives with respect to $\beta$. This is entirely
equivalent to performing an OPE-type expansion, which usually corresponds
vacuum-expanding a bi-local product operator $\hat{\mathcal{O}}$
in a ``complete basis'' of local operators \cite{Qi:2018bje}
\begin{align*}
\hat{\mathcal{O}} & =\sum_{\hat{A}}\left\langle \hat{A}^{\dagger}\mathcal{O}\right\rangle \hat{A}
\end{align*}
which will relate higher correlators to products of lower correlators
of some ``complete basis'' of operators. In this case, we take the
identity operator and the Gram-Schmidt processed Hamiltonian $\delta\hat{H}\equiv\hat{H}-\left\langle \hat{H}\right\rangle $
as two unnormalized operators of our basis. Surprisingly, (\ref{eq:OPE-like relation})
then implies that these two operators are sufficient to find the asymptotic
time-ordered four-point correlator at large $q$!

One may then wonder what is occurring when $\theta_{3}>\theta_{1}>\theta_{2}$
(\ref{eq:uncrossed OTOC}), as the fermions are still uncrossed, but
the functional form of the resultant time-ordered four-point function
changes from being a simple function of $\theta_{34}^{-}$ to one
of $2\pi-\theta_{34}^{-}$. That is, the instance of $\theta_{34}^{-}$
in $\left\langle \delta\hat{H}\hat{G}_{34}\right\rangle $ in the
OPE-like relations (\ref{eq:OPE-like relation}) must replaced by
$2\pi-\theta_{34}^{-}$ (even though the correlator $\left\langle \hat{G}_{34}\right\rangle $
remains invariant under $\theta_{34}^{-}\leftrightarrow2\pi-\theta_{34}^{-}$).
To figure out what's going on, let's write out this correlator more
explicitly
\begin{align*}
F\left(\theta_{1},\theta_{2},\theta_{3},-\theta_{3}\right) & =\frac{1}{N}\sum_{j}\Tr\left(\hat{\rho}\hat{\psi}_{j}\left(\theta_{3}\right)\hat{G}_{12}\hat{\psi}_{j}\left(\theta_{4}\right)\right),\quad\theta_{3}>\theta_{1}>\theta_{2}\\
 & =\frac{1}{N}\sum_{j}\Tr\left(\hat{\rho}\hat{G}_{12}\hat{\psi}_{j}\left(\theta_{4}\right)\hat{\psi}_{j}\left(\theta_{3}-2\pi\right)\right),\quad\theta_{3}>\theta_{1}>\theta_{2}\\
 & =\left\langle \hat{G}\left(\theta_{1},\theta_{2}\right)\hat{G}\left(\theta_{4},\theta_{3}-2\pi\right)\right\rangle ,\quad\theta_{3}>\theta_{1}>\theta_{2}
\end{align*}
where the angular separation between the two times in the second operator
is $\theta_{4}-\left(\theta_{3}-2\pi\right)=2\pi-\theta_{4}+\theta_{3}=2\pi-\theta_{34}^{-}$.
At this point, we can run through the process of inserting complete
sets of operators and we naturally arrive at expressions containing
the angular separation $2\pi-\theta_{34}^{-}$ rather than $\theta_{34}^{-}$
for when $\theta_{3}>\theta_{1}>\theta_{2}$ (\ref{eq:uncrossed OTOC}).

We should also note that this phenomenon where the time-ordered behavior
arises from ``thermal fluctuations'' does not hold solely in the
large $q$ limit. In fact, it has been shown that for any $q>2$,
these relations will apply for the near-conformal limit of large $\beta\mathcal{J}$\cite{Maldacena:2016hyu}.
Furthermore, these relations apply to \emph{any} model where the fluctuations
are governed by a Schwarzian action - a.k.a. a near-conformal limit
- and not just the $SYK$ model \cite{Maldacena:2016upp}. 

\subsection{Out-of-Time-Order Correlators and Chaos}

It will be useful to define the squeezed coordinates $\pi-\phi_{ij}^{\pm}\equiv v\left(\pi-\theta_{ij}^{\pm}\right)$
where if $\theta_{ij}^{\pm}\in\left(0,2\pi\right)$, then $\phi$
will range over the squeezed domain $\phi_{ij}^{\pm}\in\pi\left(1-v,1+v\right)$
since $0\leq v<1$. Using this we may define the quantity
\[
g\left(\phi\right)\equiv1+\frac{\pi-\phi}{2\tan\frac{\phi}{2}}
\]
which allows us to write the large $N$ behavior of the connected
out-of-time-order correlator $\theta_{1}>\theta_{3}>\theta_{2}$ on
one line
\begin{align}
\frac{\mathcal{F}}{F_{d}} & =\frac{2\tan\frac{\pi v}{2}}{\frac{\pi v}{2}+\cot\frac{\pi v}{2}}g\left(\phi_{12}^{-}\right)g\left(\phi_{34}^{-}\right)-\frac{2\sin\frac{\phi_{12}^{+}}{2}}{\cos\frac{\pi v}{2}\sin\frac{\phi_{12}^{-}}{2}\sin\frac{\phi_{34}^{-}}{2}}-\frac{\tan\frac{\pi v}{2}\left(\pi-\phi_{12}^{+}\right)}{\tan\frac{\phi_{12}^{-}}{2}\tan\frac{\phi_{34}^{-}}{2}}\label{eq:crossed OTOC}
\end{align}
where $v=v\left(\beta\mathcal{J}\right)$ is determined from (\ref{eq:determine v}).
In the limit of large $\beta\mathcal{J}$ where $v$ approaches $1$,
this agrees with the large $q$ limit of previous large $\beta\mathcal{J}$
results \cite{Maldacena:2016hyu,Maldacena:2016upp,Kitaev2018}. Unlike
previous results, this expression applies outside of the large $\beta\mathcal{J}$
regime, so long as one is at large $q$. Indeed, one may explicitly
check small $\beta\mathcal{J}$ validity by expanding the first couple
terms in the four-point function (\ref{eq:4pt defn}). In other words,
this is the first expression for a large $N$ $SYK$ four-point function
that is correct both at strong and weak coupling. Particularly, we
see that the entirety of the $\beta\mathcal{J}$ dependence has been
absorbed into the parameter $v$. Furthermore, after going to the
rescaled coordinates the four-point function only has three different
types of $v$-dependence, all of which diverge linearly as $v\rightarrow1$. 

One way of defining scrambling is when the Lorentzian evolution of
two operators in an out-of-time-order four-point correlator causes
the overall function to deviate significantly from the disconnected
correlator. In this case, since we have $F\equiv F_{d}+\mathcal{F}/N$,
scrambling can be thought of as occurring when Lorentzian evolution
causes $\mathcal{F}/F_{d}\sim N$. Now, as for the Euclidean positioning
of the operators in a scrambling setup, there are many different configurations
one can consider which preserve the crossed nature of the ordering.
Regardless, one then then Lorentzian evolves the system so that $\left(\tau_{1},\tau_{2}\right)\rightarrow\left(\tau_{1}+it,\tau_{2}+it\right)$
which makes it so $\theta_{12}^{-},\theta_{34}^{-},$ and $\theta_{34}^{+}$
are held fixed while $\theta_{12}^{+}\rightarrow\theta_{12}^{+}\pm4\pi it/\beta$.
In other words, we can consider Lorentzian evolution by giving $\phi_{12}^{+}/2$
an imaginary component $\phi_{12}^{+}/2\rightarrow\phi_{12}^{+}/2+i\frac{2\pi v}{\beta}t$.
Plugging this into (\ref{eq:crossed OTOC}), we see that there are
both linearly and exponentially growing terms in the out-of-time-order
correlator. Keeping only the exponentially growing term, we find that
\[
\frac{\mathcal{F}}{F_{d}}\sim-\frac{e^{i\mathrm{sgn}\left(t\right)\left(\pi-\phi_{12}^{+}\right)/2}e^{2\pi tv/\beta}}{\cos\frac{\pi v}{2}\sin\frac{\phi_{12}^{-}}{2}\sin\frac{\phi_{34}^{-}}{2}}+\mathcal{O}\left(2\pi vt/\beta\right)
\]
Thus, large $q$ scrambling occurs when
\begin{align*}
t_{scr} & \sim\frac{\beta}{2\pi v}\left(\log\left(N\cos\frac{\pi v}{2}\right)+\log\left(\sin\frac{\phi_{12}^{-}}{2}\sin\frac{\phi_{34}^{-}}{2}\right)\right)\\
 & \sim\frac{\beta}{2\pi v}\left(\log\left(N\cos\frac{\pi v}{2}\right)+\log\left(\cos\left(\frac{v}{2}\left(\pi-\theta_{12}^{-}\right)\right)\cos\left(\frac{v}{2}\left(\pi-\theta_{34}^{-}\right)\right)\right)\right)
\end{align*}
where we note that at large $\beta\mathcal{J}$, $\cos\frac{\pi v}{2}\rightarrow\frac{\pi}{\beta\mathcal{J}}\left(1-\frac{2}{\beta\mathcal{J}}+...\right)$,
which means that the scrambling time can be nontrivially modified
depending of whether the Euclidean angle differences $\left(\theta_{12}^{-},\theta_{34}^{-}\right)$
are comparable to $1/\beta\mathcal{J}$. 

Now since $0\leq v<1$, the Lyapunov exponent of the large $q$ $SYK$
model is always below the chaos bound \cite{Maldacena:2015waa}
\[
\lambda_{L}=\frac{2\pi}{\beta}v<\lambda_{L}^{bound}=\frac{2\pi}{\beta}
\]
It is interesting to consider the fact that the rate of growth of
the connected out-of-time-order four-point correlator is also governed
by the thermalization timescale $t_{thermal}\sim\frac{\beta}{v}$.
Alternatively stated, we have derived that the thermalization rate
and the Lyapunov exponent are the same in the large $q$ $SYK$ model
for \emph{all energies}, even when the system is described by perturbative
physics. 

\section{Discussion}

It is interesting how the functional form of the large $q$ SYK out-of-time-order
correlator (\ref{eq:crossed OTOC}) is universal: everything can be
absorbed in a new parameter $v$, which is monotonic in the coupling
and temperature (\ref{eq:determine v}). Going to low temperature/strong
coupling, one may check that it matches the previously-derived low-temperature
near-conformal expression \cite{Maldacena:2016hyu,Maldacena:2016upp,Kitaev2018}
and has therefore re-summed certain corrections sub-leading in $\beta\mathcal{J}$. 

Originally, such terms arise from the kinetic term's breaking of the
low-temperature conformal symmetry of the SYK action, which manifests
in the saddle-point equation as a derivative (or $\left[G_{0}\right]$
in Matsubara space) (\ref{eq:Orig Sad-Pt}). Now, this breaking has
not disappeared in the large $q$ limit and manifests as a constraint
on the resultant large $q$ Liouville theory that would otherwise
be conformal \cite{Maldacena:2016hyu,Maldacena:2018lmt}: the kinetic
term in the original action restricts the Liouville field variable
and its fluctuations to only those that are zero at zero time difference.
Thus, it is still true that at large $q$ one may think of the low
energy physics as being described by a spontaneously and explicitly
broken conformal symmetry. In other words, taking the large $q$ limit
preserves many interesting properties of the $SYK$ model, yet makes
the theory as a whole much more tractable.

We remind the reader of our result that the large $q$ thermalization
rate and scrambling rate (Lyapunov exponent) are the same at all couplings
and energies. This is quite significant, since the two correspond
to different physics. From perspective of operator dynamics, the thermalization
time is when an operator has appreciably changed and become \emph{anything}
else; in contrast, the scrambling time is when an operator has become
\emph{everything} else. More specifically, the process of scrambling
as defined by OTO decay refers to the increasing non-commutativity
between some evolving operator with some simple reference operator(s)
\cite{Stanford:2015owe,Shenker:2013pqa,Lashkari:2011yi}. Now, characterizing
how ``much'' an operator does not commute with some reference operator(s)
orders the space of operators into ``smaller'' and ``larger''
operators\footnote{This is similar in spirit to, but not the same as, organizing the
space of operators into closer and further operators}; it is in this sense that that the Lyapunov exponent is the rate
of ``operator growth'' \cite{Qi:2018bje,Roberts:2018mnp}. It is
thus clear that the rates of operator change (thermalization) and
growth (scrambling) need not be the same, hence why it is so interesting
that these $q$-local interactions regardless of energy or coupling
cause these two kinds of operator dynamics to occur at the same rate. 

The infinite temperature epidemic picture obtained by repeatedly applying
commutations of the Hamiltonian to a simple operator naturally produces
such a prediction \cite{Roberts:2018mnp}. The fact that the agreement
between these two rates occurs at all energies suggests that at any
energy, some form of thermally renormalized epidemic model applies
\cite{Qi:2018bje}. Furthermore, there is a growing body of evidence
that holographic OTOs are expectation values of bulk (null) momenta
for boundary-perturbed states $\hat{\mathcal{O}}\left(t\right)\ket{TFD}$
\cite{Maldacena:2017axo,Shenker:2014cwa,Maldacena:2016upp}. In other
words, the operator dynamics of holographic theories seems to be closely
related to classical mechanics on black hole geometries \cite{Brown:2018kvn,Qi:2018bje,Mousatov:2019xmc,Lin:2019qwu,Susskind:2019ddc,Lin:2019kpf}.
Having a specific understanding of how this occurs would greatly aid
in explaining the relation between semi-classical gravity and large
$N$ quantum mechanics. We are only now beginning to learn about the
smooth and simple structures that underly the operator dynamics of
theories with many internal degrees of freedom \cite{Lucas:2019cxr}.

\section*{Acknowledgements }

We are grateful to Adam Levine, Henry Lin, Andy Lucas, Juan Maldacena,
Phil Saad, Steve Shenker, Douglas Stanford, Lenny Susskind, and Ying
Zhao for useful discussions. AS is supported by the Simons Foundation.

Research at Perimeter Institute is supported in part by the Government
of Canada through the Department of Innovation, Science and Economic
Development Canada and by the Province of Ontario through the Ministry
of Economic Development, Job Creation and Trade.

\bibliographystyle{utphys}
\bibliography{TwistingCorrelators}

\appendix

\section{Twisting Correlators\label{sec:Twisting-Correlators}}

First, we temporarily purify the system. This will (often implicitly)
involve the introduction of a maximally entangled state $\ket{0}$
that satisfies
\begin{equation}
\left(A_{L}-e^{i\varphi_{A}}A_{R}\right)\ket{0}=0\label{eq:max entangled}
\end{equation}
for all ``basic'' operators $\left(A,B,etc\right)$\footnote{Note that if $\left[A_{L},B_{L}\right]_{\xi}=C_{L}$, then (\ref{eq:max entangled})
implies that $\left[A_{R},B_{R}\right]_{\xi}=-e^{i\left(\varphi_{C}-\varphi_{A}-\varphi_{B}\right)}C_{R}$}. Suppose that we want the four-point function $\left\langle ABCD\right\rangle _{\beta}$
with $\pi>\theta_{A}>\theta_{B}>\theta_{C}>0$ and $\theta_{D}=-\theta_{C}$,
then we have that $\left\langle ABCD\right\rangle _{\beta}\propto\left\langle A_{L}B_{L}C_{L}D_{R}\right\rangle _{TFD}$
where the proportionality factor is a simple phase. For ensembles
of four-point functions, we have $\sum_{CD}w_{C}w_{D}\left\langle ABCD\right\rangle _{\beta}\propto\sum_{CD}w_{C}w_{D}\left\langle A_{L}B_{L}C_{L}D_{R}\right\rangle _{TFD}$.

At this point, we have yet to do anything particularly non-trivial;
the crux is the next step. There exists choices of $\left(C,D\right)$
and weightings $\left(w_{C},w_{D}\right)$ such that the doubled operator
\[
V\propto\sum_{CD}w_{C}w_{D}C_{L}D_{R}
\]
generates twists of doubled coherent states: $\left[V,a_{L}\right]=V_{LL}a_{L}+V_{LR}a_{R}$
and $\left[V,a_{R}\right]=V_{RL}a_{L}+V_{RR}a_{R}$ where $a_{L/R}$
are annihilation operators. Consequently, $M\equiv e^{-\mu V}$ maps
one doubled coherent state into another: $M\ket{\alpha_{L},\alpha_{R}}=\ket{M_{LL}\alpha_{L}+M_{LR}\alpha_{R},M_{RL}\alpha_{L}+M_{RR}\alpha_{R}}$.
Since path integrals\footnote{Note that typically path integrals of bosonic degrees of freedom such
as $\phi\left(x\right)$ are often not often expressed in terms of
$\alpha\left(x\right)\equiv\phi\left(x\right)+i\pi\left(x\right)$,
since we usually integrate out $\pi\left(x\right)$. However, it is
not difficult to first translate the $\alpha\left(x\right)$ twist
conditions into twist conditions on $\phi\left(x\right)$ and $\pi\left(x\right)$
before attempting to integrate out $\pi\left(x\right)$. } are constructed from coherent states, the effect of $M$ is solely
to provide a twist condition on the fields. If we then calculate the
correlator $\mathcal{G}_{AB}\equiv\left\langle A_{L}B_{L}M\right\rangle _{TFD}$
with the twist conditions applying to all configurations, then we
may then obtain the four-point function via $\lim_{\mu\rightarrow0}\partial_{\mu}\mathcal{G}_{AB}=\left\langle A_{L}B_{L}V\right\rangle _{TFD}\propto\sum_{CD}w_{C}w_{D}\left\langle ABCD\right\rangle _{\beta}$.
At this point, we need not continue to work in the purified system,
as we may de-purify by sewing the left-right configuration integrals
together by defining the continuous field:
\begin{equation}
A\left(\theta\right)=\begin{cases}
A_{L}\left(\theta\right) & 0\leq\theta\leq\pi\\
-e^{i\varphi_{A}}A_{R}\left(2\pi-\theta\right) & \pi\leq\theta<2\pi
\end{cases}\label{eq:depurification conditions}
\end{equation}
This allow us to cast the doubled twist conditions in terms of the
original system. We thereby conclude that to determine (weighted averages
of) four-point functions that can be cast as $\left\langle A_{L}B_{L}V\right\rangle _{TFD}$
for $V$ which generates $LR$ coherent state twists, one may instead
calculate the $AB$ two-point function, with the restriction on the
path integral that all configurations obey de-purified twist conditions. 

Suppose we are interested in the finite temperature four-point function
involving two pairs of Hermitian fermionic operators
\begin{equation}
F\left(\theta_{1},\theta_{2},\theta_{3},\theta_{4}\right)=\frac{1}{N^{2}}\sum_{i,j=1}^{N}\left\langle \mathcal{T}\left[\psi_{i}\left(\theta_{1}\right)\psi_{i}\left(\theta_{2}\right)\psi_{j}\left(\theta_{3}\right)\psi_{j}\left(\theta_{4}\right)\right]\right\rangle _{\beta}\label{eq:4pt defn}
\end{equation}
where $\mathcal{T}$ is the Euclidean time-ordering symbol and $\theta\equiv2\pi\tau/\beta$.
Defining the coordinates $\theta_{ij}^{\pm}\equiv\theta_{i}\pm\theta_{j}$,
we use time-translation invariance to set $\theta_{34}^{+}=0$. Thus,
without loss of generality we take $\theta_{3}=-\theta_{4}=\theta_{34}^{-}/2$.
As discussed in Appendix (\ref{sec:Fundamental-Domain-of}), this
choice makes it clear that for $C\left(P\right)T$-invariant systems
it is sufficient to solve for $F_{iijj}$ in one half of the $\left(\theta_{1},\theta_{2}\right)$
domain defined by $0<\theta_{12}^{\pm}<2\pi$ and $0<\theta_{34}^{-}<2\pi$.
In other words, we need only study one-eighth of the typical fundamental
domain. Essentially, each of the operator pairings as well as the
$C\left(P\right)T$ invariance cuts the domain in half.

With the domain restricted, we need to figure out how to calculate
$F$. In our case, it will be extremely useful to temporarily purify
the calculation via the maximally entangled state $\ket{0}$\footnote{There is no unique maximally entangled state, since entanglement is
unchanged by local unitaries. Thus, let us choose the simplest state
$\ket{\tilde{0}}$ such that the left and right algebras are the same.
To guarantee this, we start by noting that if $\left(\psi_{i}^{L}+c_{i}\psi_{i}^{R}\right)\ket{\tilde{0}}=0$,
then one has that $2\delta_{ij}\ket{0}=\left\{ \psi_{i}^{L},\psi_{j}^{L}\right\} \ket{0}=-c_{i}c_{j}\left\{ \psi_{i}^{R},\psi_{j}^{R}\right\} \ket{0}$,
which gives that $c_{i}c_{j}\left\{ \psi_{i}^{R},\psi_{j}^{R}\right\} =-2\delta_{ij}$.
One is then naturally led to the choice $c_{i}=\pm i$, where in our
case we chose $c_{i}=+i$. Of course, the ambiguity in this choice
has no effect on physical predictions.}, which satisfies $\left(\psi_{i}^{L}+i\psi_{i}^{R}\right)\ket{0}=0$.
We find
\begin{align*}
F\left(\theta_{1},\theta_{2},\theta_{3},-\theta_{3}\right) & =-\frac{i}{N^{2}}\sum_{i,j=1}^{N}\ev{\mathcal{T}\left[\rho\psi_{i}^{L}\left(\theta_{1}\right)\psi_{i}^{L}\left(\theta_{2}\right)\psi_{j}^{L}\left(\theta_{3}\right)\psi_{j}^{R}\left(-\theta_{3}\right)\right]}{0}
\end{align*}
where $\mathcal{T}$ is the contour-ordering symbol and $\rho$ is
the purified Gibbs state such that $\rho^{1/2}\ket{0}=\ket{TFD}$:
$\rho=e^{-\beta\left(H_{L}+H_{R}\right)/2}/Z\left(\beta\right)$.

We then proceed to study a more complicated object
\begin{align*}
\mathcal{Z}' & \equiv\frac{1}{N}\sum_{i=1}^{N}\ev{\mathcal{T}\left[\rho\psi_{i}^{L}\left(\theta_{1}\right)\psi_{i}^{L}\left(\theta_{2}\right)M\left(\theta_{3}\right)\right]}{0}\quad M\equiv\exp\left(-\frac{i\mu}{2}\sum_{j=1}^{N}\psi_{j}^{L}\psi_{j}^{R}\right)
\end{align*}
where $M$ is of the form $M=e^{-\mu V}$with $V$ hermitian. We see
that this object generates the four-point function: $\frac{2}{N}\lim_{\mu\rightarrow0}\partial_{\mu}\mathcal{Z}'=F$.
Then, we note that it naturally factorizes into the product
\begin{gather}
\mathcal{Z}'=\mathcal{Z}\,\mathcal{G}_{LL}\left(\theta_{1},\theta_{2}\right)\nonumber \\
\mathcal{Z}\equiv\ev{\mathcal{T}\left[\rho M\left(\theta_{3}\right)\right]}{0}\qquad\mathcal{G}_{LL}\left(\theta_{1},\theta_{2}\right)\equiv\frac{\mathcal{Z}^{-1}}{N}\sum_{i=1}^{N}\ev{\mathcal{T}\left[\rho\psi_{i}^{L}\left(\theta_{1}\right)\psi_{i}^{L}\left(\theta_{2}\right)M\left(\theta_{3}\right)\right]}{0}\label{eq:twisted two-point function def}
\end{gather}
Furthermore, by taking derivatives and setting $\mu\rightarrow0$,
we see that this factorization implies that
\begin{align}
\frac{2}{N}\lim_{\mu\rightarrow0}\partial_{\mu}\mathcal{Z}' & =\frac{2}{N}\lim_{\mu\rightarrow0}\left(\partial_{\mu}\mathcal{Z}\right)\mathcal{\mathcal{G}}+\frac{2}{N}\lim_{\mu\rightarrow0}\mathcal{Z}\left(\partial_{\mu}\mathcal{G}\right)\nonumber \\
\Rightarrow F\left(\theta_{1},\theta_{2},\theta_{3},-\theta_{3}\right) & =-iG_{LL}\left(\theta_{12}^{-}\right)G_{LR}\left(\theta_{34}^{-}\right)+\frac{2}{N}\lim_{\mu\rightarrow0}\partial_{\mu}\mathcal{G}_{LL}\left(\theta_{1},\theta_{2}\right)\label{eq:four-point relation}
\end{align}
Thus, if we are able to determine the regular two-point functions
$G_{LL}$ and $G_{LR}$, as well as the twisted two-point function
$\mathcal{G}_{LL}$ to leading order in $\mu$, then we have obtained
the full four-point function. 

As discussed in \cite{Qi:2018bje}, the role of $M=\exp\left(-\frac{i\mu}{2}\sum_{j=1}^{N}\psi_{j}^{L}\psi_{j}^{R}\right)$
in (\ref{eq:twisted two-point function def}) is solely to twist the
boundary condition of the fermions from being continuous as one approaches
$\theta_{0}$ from either direction to
\begin{equation}
\lim_{\theta\rightarrow\theta_{3}^{+}}\left(\begin{array}{c}
\psi^{L}\left(\theta\right)\\
i\psi^{R}\left(\theta\right)
\end{array}\right)=\left(\begin{array}{cc}
\cosh\left(\mu\right) & -\sinh\left(\mu\right)\\
-\sinh\left(\mu\right) & \cosh\left(\mu\right)
\end{array}\right)\lim_{\theta\rightarrow\theta_{3}^{-}}\left(\begin{array}{c}
\psi^{L}\left(\theta\right)\\
i\psi^{R}\left(\theta\right)
\end{array}\right)\label{eq:Op Bdry Cond}
\end{equation}
If we then de-purify the calculation by stitching the left and right
fields together
\begin{equation}
\psi\left(\theta\right)=\begin{cases}
\psi_{L}\left(\theta\right) & 0\leq\theta\leq\pi\\
i\psi_{R}\left(2\pi-\theta\right) & \pi\leq\theta<2\pi
\end{cases}\label{eq:De-Purify}
\end{equation}
making sure carry along our boundary conditions (\ref{eq:Op Bdry Cond})
which become
\begin{align}
\lim_{\theta\rightarrow\theta_{3}^{+}}\left(\begin{array}{c}
\psi\left(\theta\right)\\
\psi\left(2\pi-\theta\right)
\end{array}\right) & =\left(\begin{array}{cc}
\cosh\left(\mu\right) & -\sinh\left(\mu\right)\\
-\sinh\left(\mu\right) & \cosh\left(\mu\right)
\end{array}\right)\lim_{\theta\rightarrow\theta_{3}^{-}}\left(\begin{array}{c}
\psi\left(\theta\right)\\
\psi\left(2\pi-\theta\right)
\end{array}\right)\label{eq:Bdry Conds}
\end{align}
then we obtain we see that our four-point function relation (\ref{eq:four-point relation})
becomes
\begin{equation}
F\left(\theta_{1},\theta_{2},\theta_{3},-\theta_{3}\right)=G\left(\theta_{12}^{-}\right)G\left(\theta_{34}^{-}\right)+\frac{2}{N}\lim_{\mu\rightarrow0}\partial_{\mu}\mathcal{G}\left(\theta_{1},\theta_{2}\right)\label{eq:Final Four-point relation}
\end{equation}
where $\mathcal{G}$ is now a two point function on a single Hilbert
space, satisfying the bi-local twist conditions (\ref{eq:Bdry Conds}).
Thus, we conclude that the factorization (\ref{eq:twisted two-point function def})
that was natural in the doubled theory is actually large $N$ factorization.

Defining 
\[
F=F_{d}+\frac{\mathcal{F}}{N}
\]
 where $F_{d}\equiv G\left(\theta_{12}^{-}\right)G\left(\theta_{34}^{-}\right)$,
then we see that the connected part of the four-point function is
given by
\[
\mathcal{F}\left(\theta_{1},\theta_{2},\theta_{3},-\theta_{3}\right)=2\lim_{\mu\rightarrow0}\partial_{\mu}\mathcal{G}\left(\theta_{1},\theta_{2}\right)
\]
Oftentimes, we are more concerned with the ratio of the connected
and disconnected four-point functions, which is given by
\[
\frac{\mathcal{F}\left(\theta_{1},\theta_{2},\theta_{3},-\theta_{3}\right)}{F_{d}}=\frac{2}{G\left(\theta_{34}^{-}\right)}\lim_{\mu\rightarrow0}\partial_{\mu}\ln\mathcal{G}\left(\theta_{1},\theta_{2}\right)
\]

For this reason, we will switch variables to $\nu\equiv\mu G\left(\theta_{34}^{-}\right)$
before taking our limits:
\[
\frac{\mathcal{F}\left(\theta_{1},\theta_{2},\theta_{3},-\theta_{3}\right)}{G\left(\theta_{12}^{-}\right)G\left(\theta_{34}^{-}\right)}=2\lim_{\nu\rightarrow0}\partial_{\nu}\ln\mathcal{G}\left(\theta_{1},\theta_{2}\right)
\]

\section{Fundamental Domain of Four-Point Functions\label{sec:Fundamental-Domain-of}}

A four-point function of \textit{two pairs} of operators has much
more symmetry than a generic four-point function, which will we use
to vastly restrict the fundamental domain. 

First, we note that $F_{iijj}$ is odd when the first pair of times
$\left(\tau_{1},\tau_{2}\right)$ are reflected across the lines $\tau_{1}-\tau_{2}=n\beta\quad\forall n\in\mathbb{Z}$
\begin{equation}
F_{iijj}\left(\tau_{2}+n\beta,\tau_{1}-n\beta,\tau_{0}\right)=-F_{iijj}\left(\tau_{1},\tau_{2},\tau_{0}\right)\label{eq:Minus Reflection}
\end{equation}
Thus, knowledge of the four-point function for $0<\tau_{1}-\tau_{2}<\beta$
is sufficient to determine it everywhere. A similar condition on $\tau_{3}$
and $\tau_{4}$ implies that we may use as a fundamental domain $0<\tau_{3}-\tau_{4}<\beta$,
which for our parametrization implies that we need only understand
$F_{iijj}\left(\tau_{1},\tau_{2},\tau_{0}\right)$ for $0<\tau_{0}<\beta/2$
to determine the four-point function everywhere.

Second, $C\left(P\right)T$ invariance implies that $F_{iijj}\left(\tau_{1},\tau_{2},\tau_{3},\tau_{4}\right)$
satisfies
\begin{align*}
F_{iijj}\left(\tau_{1},\tau_{2},\tau_{3},\tau_{4}\right) & =F_{jjii}\left(\beta-\tau_{4},\beta-\tau_{3},\beta-\tau_{2},\beta-\tau_{1}\right)\\
 & =F_{iijj}\left(\beta-\tau_{1},\beta-\tau_{2},\beta-\tau_{3},\beta-\tau_{4}\right)\\
 & =F_{iijj}\left(\beta-\tau_{2},\beta-\tau_{1},\beta-\tau_{4},\beta-\tau_{3}\right)
\end{align*}
 Now we see that with $\hat{G}_{12}$ our choice $\tau_{3}=\beta-\tau_{0}$,
$\tau_{4}=\tau_{0}$, we have the relationship
\begin{equation}
F_{iijj}\left(\tau_{1},\tau_{2},\tau_{0}\right)=F_{iijj}\left(\beta-\tau_{2},\beta-\tau_{1},\tau_{0}\right)=F\left(\beta-\tau_{1},-\tau_{2},\tau_{0}\right)\label{eq:Plus Reflection}
\end{equation}
Combined with anti-periodicity of $F_{iijj}$ under $\beta$ translations
of $\tau_{1}$ and $\tau_{2}$, we conclude that $F_{iijj}$ is even
when $\left(\tau_{1},\tau_{2}\right)$ are reflected across the lines
$\tau_{1}+\tau_{2}=m\beta\quad\forall m\in\mathbb{Z}$. Consequently,
we may restrict our domain of $F\left(\tau_{1},\tau_{2},\tau_{0}\right)$
to $0<\tau_{1}+\tau_{2}<\beta$. In fact, we may actually restrict
this further, since we note that composing the reflections and translations
produces non-trivial point reflections. Specifically, the second equality
in (\ref{eq:Plus Reflection}) tells us that $F$ is even under reflections
through the point $\left(\tau_{1}=\beta/2,\tau_{2}=0\right)$. We
thereby conclude that it is sufficient to solve for $F_{iijj}$ in
one half of the $\left(\tau_{1},\tau_{2}\right)$ domain defined by
$0<\tau_{1}\pm\tau_{2}<\beta$ as well as for $\tau_{3}=\beta-\tau_{0}$,
$\tau_{4}=\tau_{0}$, with $\tau_{0}<\beta/2$ to determine $F_{iijj}$
everywhere. 

Note that the fundamental domain for two pairs of operators is only
one-eighth of the fundamental domain for a generic four-point functions.
This factor came from a halving of the fundamental domain due to each
of the swap symmetries, and a final halving due to the $C\left(P\right)T$
symmetry.
\end{document}